\documentclass[aps,prl,twocolumn,showpacs,groupedaddress,amsmath,amssymb]{revtex4}
\usepackage{graphicx}
\usepackage{color}
\usepackage{bm}
\usepackage{latexsym}

\begin{document}

\title{Experimental Observation of Stabilization and Lasing Death via Asymmetric Gain}
\author{M. Chitsazi, S. Factor, J. Schindler, H. Ramezani, F. M. Ellis, T. Kottos}
\affiliation{Department of Physics, Wesleyan University, Middletown, Connecticut 06459, USA}


\begin{abstract}
Using a pair of coupled LRC cavities we experimentally demonstrate that instabilities and amplification action 
can be tamed by a spatially inhomogenous gain. Specifically we observe the counter-intuitive phenomenon of
stabilization of the system even when the overall gain provided is increased. This behavior is directly related 
to lasing death via asymmetric pumping, recently proposed in [M. Liertzer {\it et al}., Phys. Rev. Lett. {\bf 108}, 
173901 (2012)]. The stability analysis of other simple systems reveals the universal nature of the lasing death 
phenomenon.
\end{abstract}


\maketitle

A laser oscillator consists of a gain medium embedded in an optical cavity. When the pumping level exceeds a 
threshold value that balances light leakage out of the cavity and other losses, the system self-organizes to emit a 
narrow-band coherent electromagnetic radiation \cite{H86}. Depending on the various geometric characteristics of 
the confined cavity (chaotic or integrable) \cite{shape} and the properties of the index of refraction (uniform, periodic, 
aperiodic or random) \cite{aperiodic,random} etc., various type of lasing modes have been identified and analyzed 
thoroughly during the past years. There are also laser systems in which the spatial distribution of the gain medium 
plays a significant role in determining the properties of the lasing mode \cite{T80,KS72}. Within this framework of 
spatially inhomogeneous gain, the counter-intuitive phenomenon of {\it lasing death}, i.e. the possibility to "shut 
down" a laser while the overall pump power provided to the system is increased, has been predicted in Ref. \cite{LGCSTR12} 
based on simulations that make use of a steady-state ab initio laser theory \cite{stone}. 

In this paper we experimentally demonstrate the phenomenon of lasing death (LD) using a system of
two coupled LRC circuits with active conductances $R^{-1}$ of either sign that determine gain parameters 
$\gamma_1$ and $\gamma_2$. We show that for specific pumping paths, the system undergoes a transition from stable to 
unstable dynamics and then back to stable dynamics despite the fact that the total gain $\gamma=\gamma_1+
\gamma_2$ continually increases along the path. These experimental results are theoretically understood via a stability 
analysis of several simple systems that illustrate the universal aspects of LD phenomenon.

\begin{figure}[h]
\includegraphics[width=0.85\columnwidth,keepaspectratio,clip]{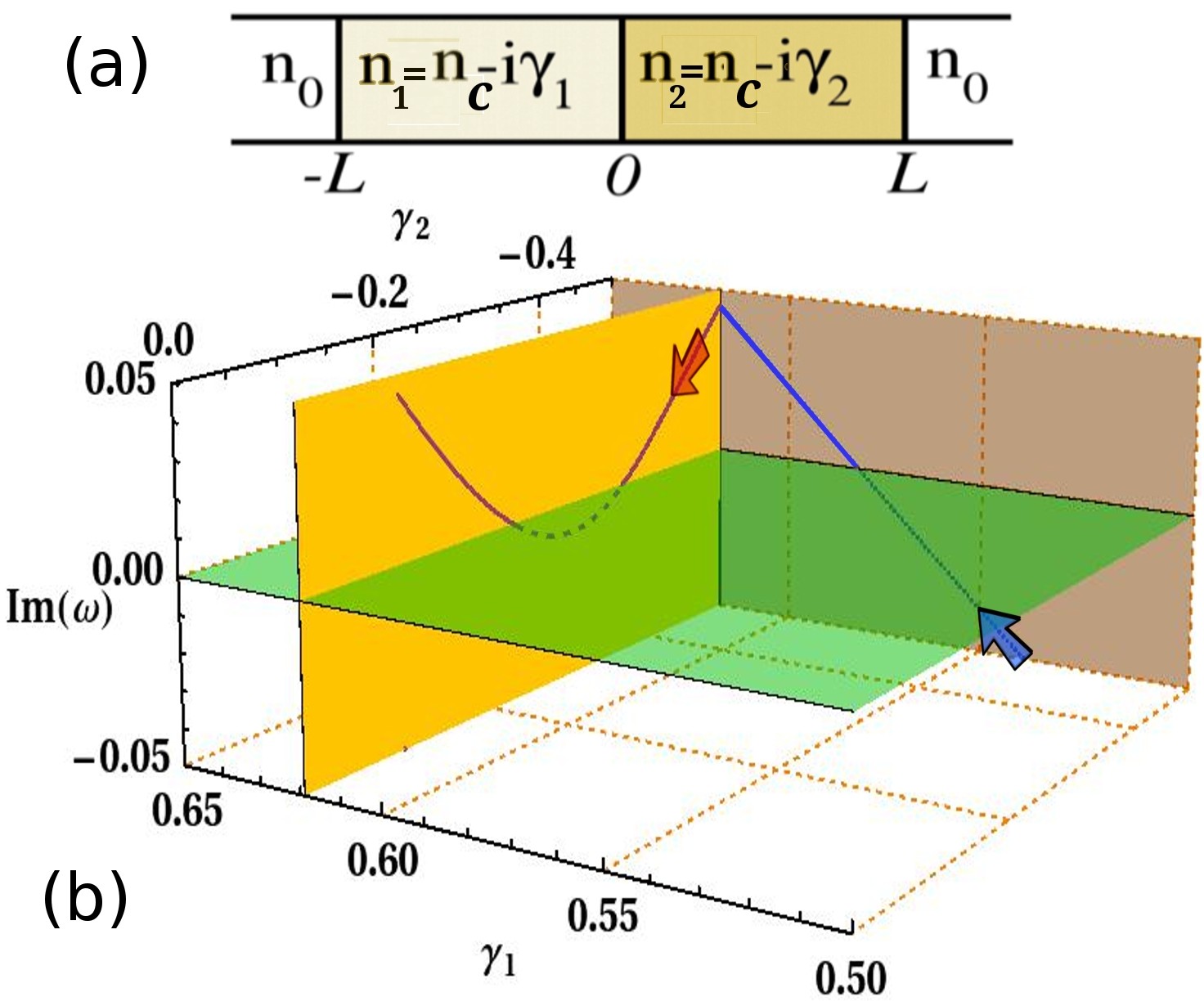}
\caption{(color online) (a) A 1D laser cavity with spatially inhomogenous gain profile. The cavity occupies the region
$-L\leq x \leq L$. The gain profile is defined by the imaginary part of the index of refraction which take the values
${\cal I}m[n_1(-L\leq x \leq 0)]=\gamma_1$ and ${\cal I}m[n_2(0\leq x \leq L)]=\gamma_2$ where $\gamma_1\neq 
\gamma2$. (b) Parametric evolution of the dominant pole \cite{gain} of the scattering matrix $S$ as the gain $\gamma_{1,2}$ 
changes}
\label{fig1} 
\end{figure}

We start our analysis by reviewing the lasing death phenomenon in the optics framework \cite{LGCSTR12}. We consider 
a $1D$ photonic cavity of length $2L$ comprised of two active regions with spatially inhomogeneous gain (see Fig. 
\ref{fig1}(a)). In the left region $-L\leq x<0 $ the index of refraction is given by $n_1=n_c-i \gamma_1$ while in the 
right region $0< x \leq L$ the index profile is given by $n_2=n_c-i\gamma_2$. $n_c$ is the real index of refraction 
associated with the cavity while $n_0$ is the background index of refraction where the cavity is embedded. The 
imaginary parts $\gamma_1,\gamma_2$ describe the gain in each region of the cavity. 

Above the lasing threshold, 
the system in general has to be treated as nonlinear \cite{stone}, as long as the amplitude is small, 
it satisfies the linear Maxwell equations with a negative (amplifying) imaginary part of the refractive index generated by 
the population inversion due to the pump \cite{H86}. Furthermore, it can be rigorously shown within semiclassical laser 
theory that the first lasing mode in any cavity is an eigenvector of the electromagnetic scattering matrix ($S$ -matrix) 
with an infinite eigenvalue; i.e., lasing occurs when a pole of the $S$-matrix is pulled "up" to the real axis by including 
gain as a negative imaginary part of the refractive index \cite{F00}. We therefore proceed to evaluate the $S-$matrix 
associated with the structure of Fig. \ref{fig1}(a).

In the arrangement of Fig. \ref{fig1}(a), a time-harmonic electric field of frequency $\omega$ obeys the Helmholtz equation:
\begin{equation}
\label{Helmholtz}
{\partial^2 E (z)\over \partial z^2} + {\omega^2 \over c^2} n^2(z) E(z) = 0\,\,\,.
\end{equation}
Eq. (\ref{Helmholtz}) admits the solution $E_0^{-}(z)=E_{f}^- \exp(ikz) +E_{b}^- \exp(-ikz)$ for 
$z<-L$ and $E_0^{+}(z)=E_{f}^+ \exp(ikz) + E_{b}^+ \exp(-ikz)$ for $z>L$ where the wavevector $k= n_0\omega/c$. 
The amplitudes of the forward and backward propagating waves outside of the cavity domain are related through the 
transfer matrix $M$:
\begin{equation}
\label{transfer}
\left(\begin{array}{c}
E_{f}^+\\
E_{b}^+
\end{array}\right)=
\left(\begin{array}{cc}
M_{11}&M_{12}\\
M_{21}&M_{22}
\end{array}\right)
\left(\begin{array}{c}
E_{f}^-\\
E_{b}^-
\end{array}\right)
\end{equation}
The transmission and reflection amplitudes for left (L) and right (R) incidence waves, are obtained from the boundary
conditions $E_b^+=0$ ($E_f^-=0$) respectively, and are defined as $t_L\equiv {E_f^{+}\over E_f^{-}}$, $r_L\equiv 
{E_b^{-}\over E_f^{-}}$; ($t_R\equiv {E_b^{-}\over E_b^{+}}$; $r_R\equiv {E_f^{+}\over E_b^{+}}$). These are expressed 
in terms of the $M-$matrix elements:
\begin{equation}
\label{trcoefficients}
t_L=t_R=t={1\over M_{22}}\,\,;\,\, r_L= -{M_{21}\over M_{22}}\,\,; \,\, r_R = {M_{12}\over M_{22}}.
\end{equation}
From these relations we evaluate the $S$-matrix as
\begin{equation}
\label{Smatrix1}
\left(\begin{array}{c}
E_{f}^+\\
E_{b}^-
\end{array}\right)=S
\left(\begin{array}{c}
E_{b}^+\\
E_{f}^-
\end{array}\right)
;\quad S=
\left(\begin{array}{cc}
r_L&t\\
t&r_R
\end{array}\right)
\end{equation}
which below the first lasing threshold, connects the outgoing wave amplitudes to their incoming counterparts. 

Scattering resonances correspond to purely outgoing boundary conditions on the $S$ matrix, i.e. $(E_f^+, E_b^-)^T\neq 0$ 
while $(E_b^+, E_f^-)^T=0$, and occur when the $S$ matrix has a pole in the complex plane. It follows from Eqs. 
(\ref{trcoefficients},\ref{Smatrix1}) that the poles $\omega_p=\omega_R-i\omega_I$ of the $S$-matrix can be identified with 
the complex zeros of the secular equation $\det (M_{22})=0$. Poles precisely on the real axis do correspond to physical states, 
and we give them the different name of \textquotedblleft threshold lasing modes\textquotedblright\ to distinguish them from 
other resonances.

In Fig. \ref{fig1}(b) we report the evolution of ${\cal I}m(\omega)$ for the dominant pole of the $S$-matrix for a specific 
path of the pumping parameters $\gamma_1,\gamma_2$ associated with the left/right portion of the cavity. The dominant 
pole is the only one to experience the stability-instability transitions within the frequency range assumed for the $\gamma_1$ 
and $\gamma_2$ gain curves \cite{gain}. First, the left part of the cavity is pumped ($0<\gamma_1<\gamma_1^{\rm max}$) 
until a pole crosses the real axis $({\cal I}m(\omega)=0)$ at some critical gain $\gamma_1^*<\gamma_1^{\rm max}$. At 
this point, a lasing state in the cavity is created. For $\gamma_1^*<\gamma_1<\gamma_1^{\rm max}$ the pole continues 
to travel upwards in the positive $({\cal I}m(\omega)>0)$ plane indicating unstable dynamics. In this regime, any physical 
system ultimately becomes non-linear and the scattering approach fails. However, we may infer from our low amplitude linear analysis the presence or absence of the lasing instability. The pumping on the left partition is now kept constant ($\gamma_1=
\gamma_1^{\rm max}$), while additional pumping, via $\gamma_2$ is applied to the right partition of the cavity. Surprisingly, this results in 
reversing the evolution of the pole back towards ${\cal I}m(\omega)=0$. At some critical value $\gamma_2=\gamma_2^*$ 
the pole re-crosses the real axis returning to ${\cal I}m(\omega)<0$. Such transitions indicate that the system returns to 
stability, i.e. the laser shuts off despite the fact that the overall pump power provided to the system has been increased.  
Further increase of $\gamma_2$ once again reverses the direction of the motion of the pole which moves upwards and 
crosses into ${\cal I}m(\omega)>0$. At this new critical value, $\gamma=\gamma_2^{**}$, 
the mode once again becomes unstable signifying a second turn on of the laser. 

\begin{figure}
\includegraphics[width=0.85\columnwidth,keepaspectratio,clip]{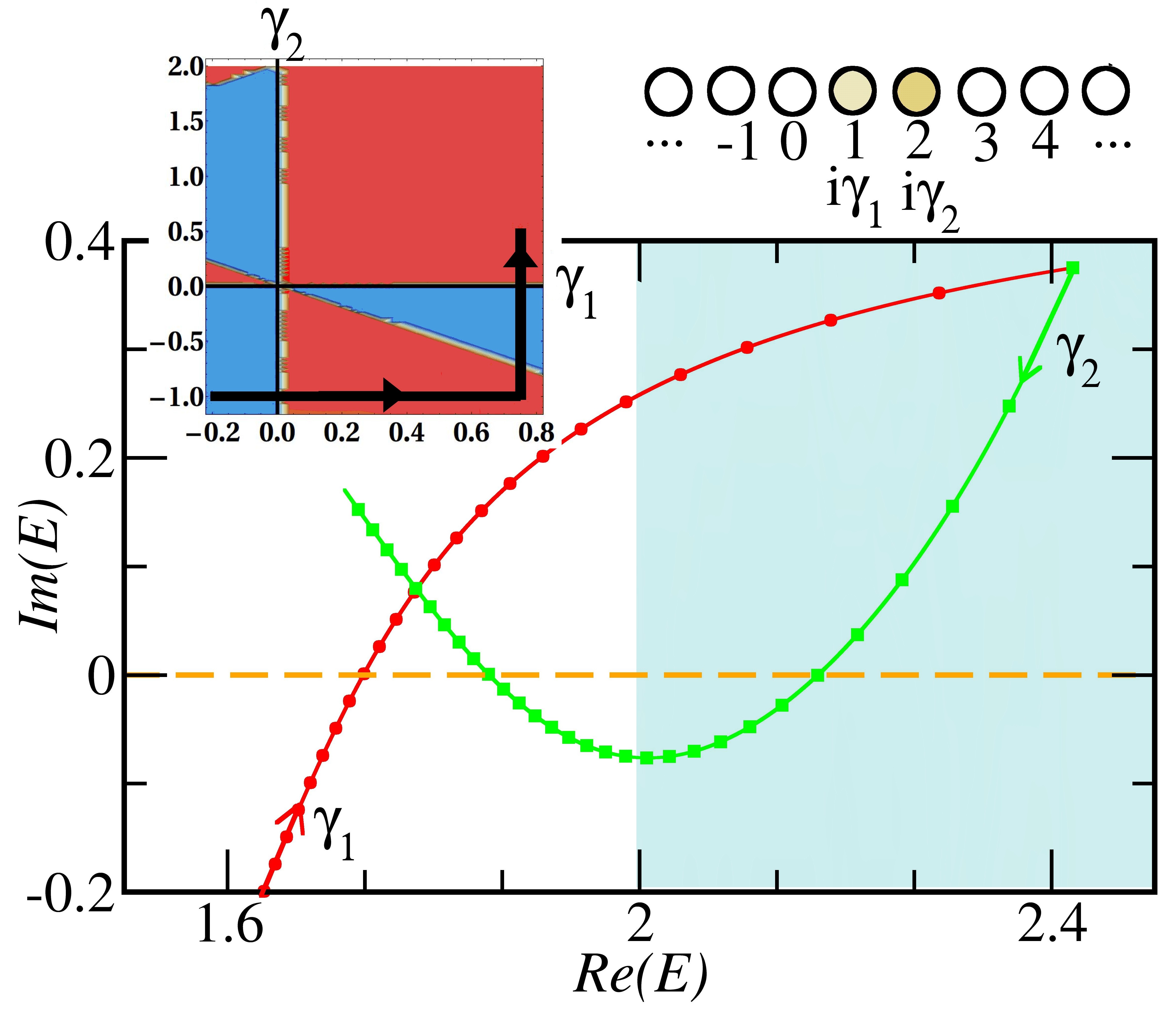}
 \caption{(color online) Right Inset: A simple tight-binding model consisting of two active discrete elements 
coupled to two semi-infinite left and right leads. Left Inset: The stability phase diagram of the TB system. Red 
areas indicate unstable $\gamma_1-\gamma_2$ domains while blue areas indicate stable domains. The arrows indicate 
the direction of the pumping path. Main panel: Parametric evolution of one of the two resonances as the pumping 
$\gamma_1$ at site $n_N=1$ and $\gamma_2$ at site $n_N=2$ changes. The arrows indicate the evolution of resonances 
as $\gamma_1, \gamma_2$ increases. The parametric evolution of the second resonance is not shown since its motion 
is symmetric to the one reported here with respect to the origin of the axis ${\cal R}e ({\cal E})$. 
}
\label{fig2}
\end{figure}

The universality of lasing death via asymmetric pumping can also be illustrated within a simple two-level model (dimer) coupled
to two leads. The 
system is shown in the inset of Fig. \ref{fig2}. The Hamiltonians of the dimer $H_d$ and of the leads $H_{\rm leads}$ read: 
\begin{eqnarray}
\label{TB}
H_d &=& \sum_{n_N=1,2} \epsilon_{n_N} |n_N\rangle\langle n_N| + (|n_N\rangle\langle n_{N}+1|+c.c.)\nonumber\\
H_{\rm leads}^{L,R} &=&\sum_{n=n_L,n_R} \epsilon_{n} |n\rangle\langle n| +  (|n\rangle\langle n+1|+c.c.)
\end{eqnarray}
where $n_L=0,\cdots, -\infty$, $n_R=3,\cdots, \infty$ and $\{|n\rangle\}$ is the Wannier basis of the TB Hamiltonian. The 
on-site potentials are $\epsilon_n=V_n+i\gamma_n$ with $\gamma_n=0$, for $n\neq1,2$ \cite{tbgain}. Furthermore, without 
loss of generality we will assume that $V_n=0$ for all $n$. The complex zeros ${\cal E}$ of the secular equation 
$\det[M_{22}({\cal E})]=0$ can be calculated analytically:
\begin{equation}
\label{EM}
 {\cal E}=\frac{i[\gamma_1 \gamma_2(\gamma_1+\gamma_2)\pm
(2+\gamma_1 \gamma_2)\sqrt{(\gamma_1-\gamma_2)^2-4}]}{2(1+\gamma_1 \gamma_2)}
\end{equation}
 In Fig. \ref{fig2} we present a parametric evolution of the poles ${\cal E}$ for a pumping path (see left inset) analogous to 
the previous discussion: an initial increase of 
$\gamma_1$ until the lasing threshold is reached, followed by an increase of $\gamma_2$. As before, during the second 
section of the path the system is first driven back towards stability (lasing death) while later on returning to instability at a second lasing threshold. We note that the pumping path within the shaded region of Fig. \ref{fig2} has to be excluded from our analysis. Here, the poles have ${\cal R}e ({\cal E})>2$ and the scattering modes fall outside of the propagation band $E(k)=2\cos(k)$ of the leads.

The generality of the lasing death phenomenon calls for a simple argument for its explanation. Using standard 
methods \cite{MW69} we write the scattering matrix elements in the form \cite{SOKG00,KW02}
\begin{equation}
\label{smatrix}
S_{\alpha,\beta}(E) = \delta_{\alpha,\beta}-i \sqrt{4-E^2}\, {\cal W}_{\alpha}^{\,T} (E-{\cal H}_{\rm eff})^{-1} {\cal W}_{\beta} \ ,
\end{equation}
where $\alpha,\beta=1,2$ and ${\cal H}_{\rm eff}$ is a $2\times 2$ effective non-hermitian Hamiltonian 
given by
\begin{equation}
\label{Heff}
{\mathcal{H}}_{\rm eff}(E)=H_d+ \Sigma(E) \sum_{\alpha} {\cal W}_{\alpha}\bigotimes {\cal W}_{\alpha}^{\,T} 
\end{equation}
The two-dimensional vectors $W_{1}=\delta_{\alpha,1}$ and $W_2=\delta_{\alpha,2}$ describe at which site we couple 
the leads with our sample while $\Sigma(E)=\frac{E-i\sqrt{4-E^2}}{2}$ is the so-called self-energy.

The poles of the $S$ matrix are equal to the complex zeros ${\cal E}$ of the following secular equation
\begin{equation}
\label{poleseq}
\det [{\cal E}-H_{\rm eff}({\cal E})]=0.
\end{equation}
Solving Eq. (\ref{poleseq}) is (in general) a difficult task. However, there are circumstances (see for 
example the LRC circuit below), for which one can neglect the dependence of $H_{\rm eff}$ on energy. In such cases 
the second term in Eq. (\ref{Heff}) 
results in a simple constant shift of the on-site potential of the Hamiltonian $H_d$.

From the above discussion we conclude that the stability of our system and the lasing death phenomenon are directly linked 
with the stability diagram and the parametric evolution of the complex eigenvalues ${\cal E}$ of $H_{\rm eff}$. Specifically 
the sign of their imaginary part ${\cal I}m({\cal E})$ defines the stability (${\cal I}m({\cal E})<0$) or instability (${\cal I}m
({\cal E})>0$) of the associated modes and therefore of the system itself. For example for the system of Eq. (\ref{TB}) we find 
that the corresponding eigenvalues 
of $H_{\rm eff}$ are given by Eq. (\ref{EM}). We include in the left inset of Fig. \ref{fig2} the associated stability diagram. 
Unstable domains are indicated with red and are associated with an imaginary part of one of the eigenvalues being larger than 
zero. The stable domains are indicated with blue and they correspond to gain parameters $\gamma_1,\gamma_2$ for which 
${\cal I}m ({\cal E})<0$. It is therefore obvious that one can select pumping paths in the $(\gamma_1,\gamma_2)$ 
plane which lead to transitions from stability to instability and back to stability while $\gamma_1+\gamma_2$ is continuously 
increasing through the pumping process.

\begin{figure}[h]
\includegraphics[width=0.85\columnwidth,keepaspectratio,clip]{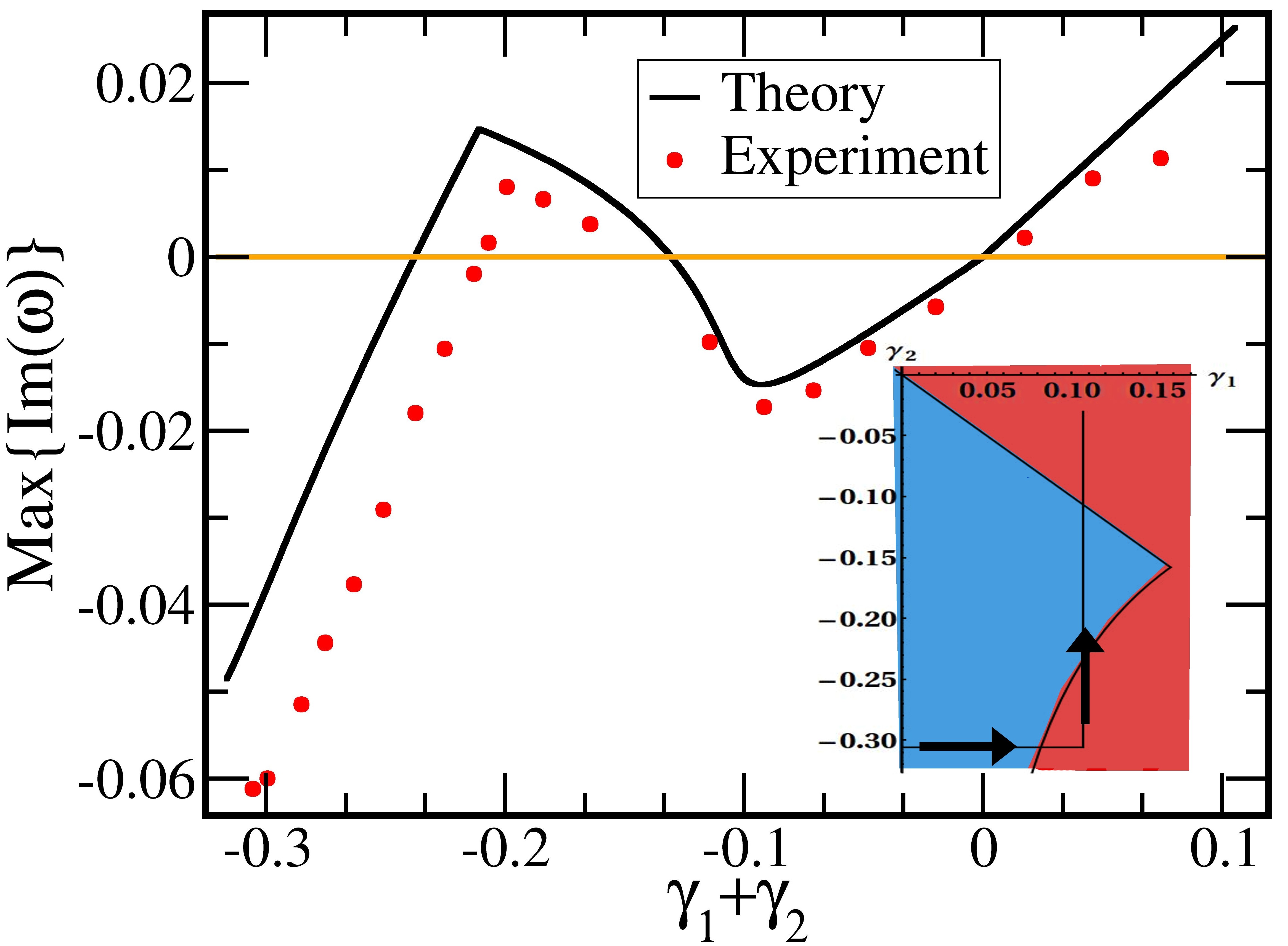}
\caption{(color online) The theoretical and experimental maximum imaginary part of all eigenfrequencies is shown as gain is 
increased along the pumping path illustrated by the stability diagram (inset) of the coupled LRC circuit with asymmetric gain 
profile. The ``lasing death'' phenomenon occurs when the path traverses a protrusion in the stability map -- most negative 
imaginary part momentarily reenters the positive region. The horizontal axis indicates total gain.
}
\label{fig3} 
\end{figure}

We experimentally demonstrate the universality of the lasing death phenomenon by exploiting its simplicity within the framework 
of electronics. To this end we consider a pair of coupled LRC circuits with asymmetric conductances of either sign responsible 
for gain parameters $\gamma_1=-\frac{1}{R_1}\sqrt{\frac{L}{C}}$ and $\gamma_2=-\frac{1}{R_2}\sqrt{\frac{L}{C}}$. The LC dimer 
heart of the circuit is identical to that used in a previous work \cite{SLLREK12}. The conductances of each LC unit are 
implemented with a positive conductance in parallel with negative impedance converters containing ``vactrols'' \cite{VCTRL}. The 
vactrols, consisting of LED controlled photo-resistors, then allow digital control of the total conductance over both positive and negative values, 
and thus exploration of the $(\gamma_1,\gamma_2)$ parameter space in all quadrants. We include only capacitive coupling and 
consider a balanced dimer with matching $LC$ resonators.

Application of the first and second Kirchhoff’s law, leads to the following set of equations for the inductor voltage $V_n$ and 
current $I_n$ ($n=1,2$)
\begin{equation}
\label{KL}
V_n=L \dot{I}_n;  \quad \frac{V_n}{R_n}+C \dot{V_n}+I_n + C_c (\dot{V}_n - \dot{V}_{3-n}) = 0
\end{equation}
In our experiment, $L=10 \mu H$ and $C=328 pF$ based on the uncoupled frequency of $\omega_1 /{2\pi} = 2.78 MHz$. The 
frequency of the second oscillator was adjusted to within $1\%$, and the capacitance coupling was $C_c=56 pF$. The dimer can 
additionally be considered as coupled to leads if the resistances $R_1$ and $R_2$ are interpreted as including the parallel 
impedance $Z_0$ of TEM transmission line leads. With $Z_0$ real and frequency independent, any non-zero voltage (1 or 2) 
is then coupled to power radiated into the corresponding transmission line.

The imaginary part of the eigenfrequencies $(e^{-i \omega t})$ of the system Eq. (\ref{KL}) dictate the stability of 
the system: the circuit is unstable if any of the modes 
have a positive imaginary part, otherwise it is stable. For a stable circuit all transient solutions are decaying, while for the unstable 
circuit, there is at least one exponentially growing solution corresponding to electronic analog of lasing into the transmission lines. 
The linear equations of motion will remain valid provided all voltages and currents remain below any nonlinearities of the actual circuit. 

Figure \ref{fig3} shows the evolution of experimental values for $Im(\omega)$ as a function of total gain, defined as $\gamma_1+
\gamma_2$ obtained along the path in the $(\gamma_1,\gamma_2)$ stability map shown in the inset. The color scheme used is 
the same as that of Fig. \ref{fig2}.  Experimental frequencies are obtained by imposing an initial current in side 1 through a forward biased 1N914 signal diode, then rapidly switching to a reverse-biased state, where the remaining contribution to the circuit is a small $\approx 0.3 pF$ junction capacitance. The subsequent voltage evolution is recorded and fit to a generic double resonance transient 
to obtain the most positive imaginary part of the eigenfrequencies. In the scattering configuration 
with TEM transmission lines of impedance $Z_0$ attached to the $LC$ nodes, these frequencies are identical to the poles of the
circuit $S$-matrix with the gain parameters $\gamma_1$ and $\gamma_2$ reduced by $\Delta\gamma =\frac{1}{Z_0}\sqrt{\frac{L}{C}}$, 
the transmission line contact loss.

Starting along the horizontal path of Fig. \ref{fig3} (inset), the "electronic lasing" turns on at a first threshold of stability as $\gamma_1$ 
is increased for fixed $\gamma_2$. Continuing along the vertical path, as $\gamma_2$ is increased with $\gamma_1$ fixed, the lasing 
death (amplification death) for our system is seen as the path cuts through a protruding section of the stability map. The imaginary part 
of the eigenfrequency becomes negative (stable) then turns around, and finally becomes permanently positive (unstable). Note that this 
occurs along a path of uniformly increasing gain. 

\begin{figure}[h]
\includegraphics[width=0.85\columnwidth,keepaspectratio,clip]{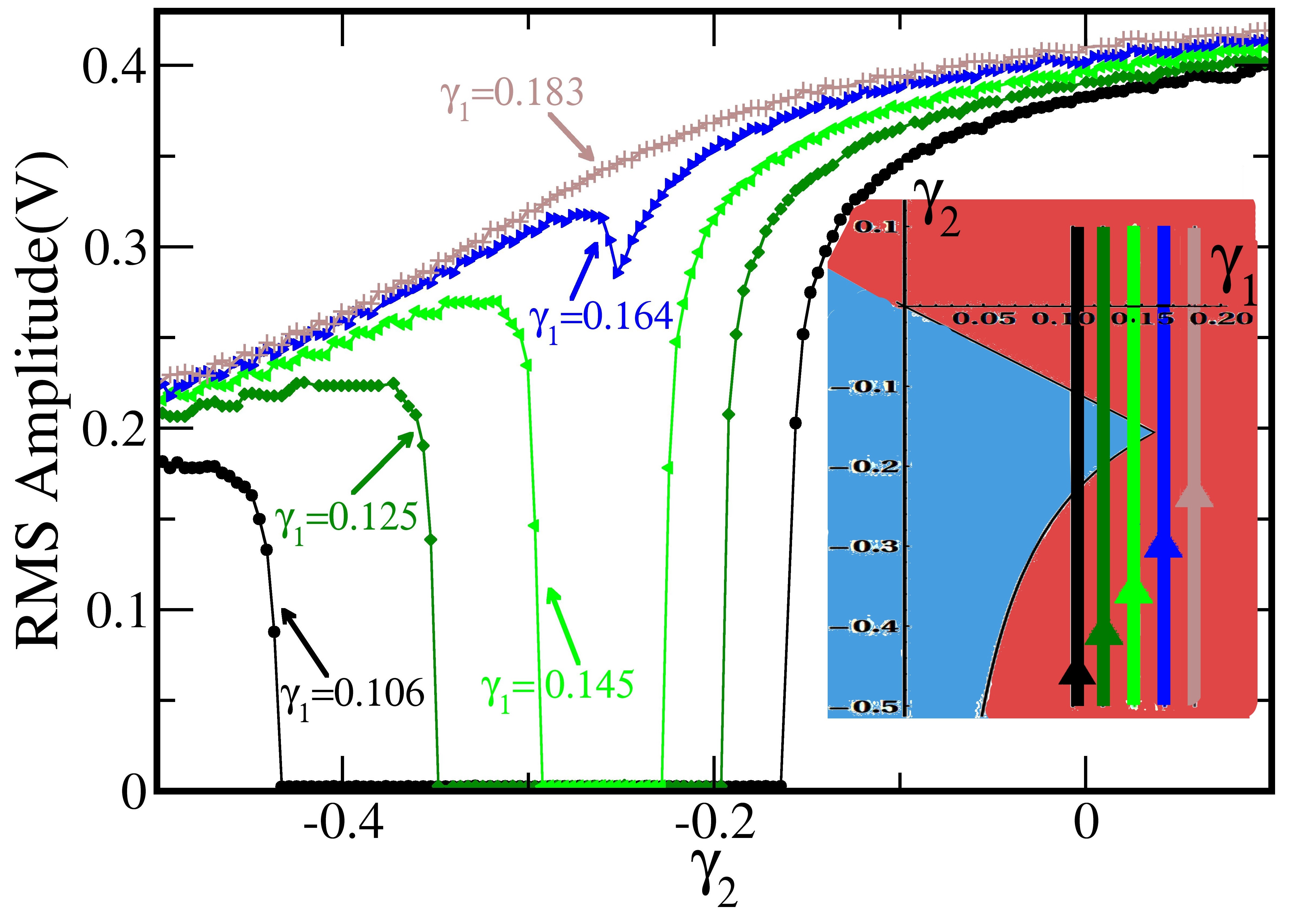}
\caption{(color online) Experimental steady-state voltage for paths of increasing $\gamma_2$ at fixed $\gamma_1$ shown in 
the inset. The voltage was measured on the side-1 LC node. Note that the width of the ``lasing death'' response diminishes as 
the overall gain increases. 
}
\label{fig4} 
\end{figure}

In a closer analog to a laser, the above behavior is also experimentally confirmed via steady state measurements of the $RMS$ 
voltage $V_{\rm rms}$ of the $n=1$ LC node as the re-entrant section of the stability diagram is traversed. When the system is 
stable, $V_{\rm rms}=0$. When the system is 
unstable, a nonzero value of $V_{\rm rms}$ is measured that is dictated by circuit saturation dynamics, which in our system was 
imposed by back-to-back 1N914 signal diodes in parallel with each LC circuit. The data in Fig. \ref{fig4} are for several paths of increasing $\gamma_2$, along the vertical lines shown in the inset, with $\gamma_1=\text{const}$. The top data set shows 
that for a large enough $\gamma_1$, the path completely misses the stable region.

To summarize, we have experimentally demonstrated the phenomenon 
responsible for lasing death in a simple system. Using two coupled LC oscillators 
with independently controlled gain, we have shown how the output can be 
extinguished over an interval of increasing system gain. This unorthodox and robust effect 
of stabilization via gain is universal and has its roots in the complicated structure 
of the stability phase diagrams that non-uniformly active structures can exhibit.

{\it Acknowledgments --} 
This research was supported by an AFOSR grant No. FA 9550-10-1-0433, and by an NSF ECCS-1128571 grant. 
SF acknowledges support from Wesleyan Faculty/Student Internship grants.


\begin{thebibliography}{99}

\bibitem{H86} H. Haken, {\it Laser Light Dynamics} (North Holland, Amsterdam, 1986).

\bibitem{shape} A. D. Stone, Nature {\bf 465}, 696 (2010); H. G. L. Schwefel, H. E. Tureci, A. Douglas Stone, R. K Chang, in 
{\it Optical Processes in Microcavities} (World Scientific 2003); J. U. N\"ockel and A. D. Stone, Nature 385, 45 (1997) 

\bibitem{aperiodic} L. Dal Negro, S. V. Boriskina, Laser \& Photonics Reviews {\bf 6}, 178 (2012)

\bibitem{random} H. Cao, Journal of Phys. A: Math. Gen. {\bf 38}, 10497 (2005).

\bibitem{T80} G. H. B. Thompson, {\it Physics of Semiconductor Laser Devices}, (Wiley \& Sons, New York, 1980), Chap. 6.

\bibitem{KS72}H. Kogelnik, C. V. Shank, J. Appl. Phys. {\bf 43}, 2327 (1972); Y. Luo, Y. Nakano, K. Tada, Appl. Phys. Lett. {\bf 56}, 1620 (1990).

\bibitem{LGCSTR12} M. Liertzer, Li Ge, A. Cerjan, A. D. Stone, H. E. T\"ureci, S. Rotter, Phys. Rev. Lett. {\bf 108}, 173901 (2012).

\bibitem{stone} H. E. T\"ureci, A. D. Stone, L. Ge, S. Rotter, and R. J. Tandy, Nonlinearity \textbf{22}, C1 (2009).


\bibitem{F00} K. M. Frahm \textit{el al.}, Europhys. Lett. \textbf{49}, 48 (2000).

\bibitem{gain}We have limited the frequency dependence of the gain profiles to exclude gain for all higher frequency poles than that shown in the figure. Furthermore we have checked numerically that all lower poles remain stable and are not presented in Fig. \ref{fig1}. 

\bibitem{tbgain} Unlike the optical cavity, ${\cal I}m (\epsilon_n)>0$ represents gain while ${\cal I}m(\epsilon_n)<0$ 
represents loss.


\bibitem{MW69} C. Mahaux and H. A Weidenm\"uller, {\it Shell Model Approach in Nuclear
Reactions}, (North-Holland, Amsterdam), (1969); I. Rotter, Rep.  Prog. Phys. {\bf 54},
635 (1991).

\bibitem{SOKG00} J. A. Mendez-Bermudez, T. Kottos, Phys. Rev. B {\bf 72}, 064108 (2005).

\bibitem{KW02} T. Kottos, M. Weiss,  Phys. Rev. Lett., {\bf 89}, 056401 (2002).

\bibitem{SLLREK12}J. Schindler, Z. Lin, J. M. Lee, H. Ramezani, F. M. Ellis, T. Kottos, J. Phys. A: Math. Theor {\bf 45},
444029 (2012).


\bibitem{VCTRL} Silonex NSL-32.

\end{thebibliography}
\end{document}